\documentclass[11pt, a4paper,onecolumn]{revtex4-2}
\usepackage[utf8]{inputenc}

\usepackage{bbm}
\usepackage{lmodern}
\usepackage{enumitem}
\usepackage{hyperref}

\usepackage{siunitx}
\usepackage{amsmath}
\usepackage{amssymb}
\usepackage{mathtools}
\usepackage{graphicx}
\usepackage{caption}
\usepackage{subcaption}
\usepackage[dvipsnames]{xcolor}

\newcommand{\braket}[2]{\left\langle\,#1\,|\,#2\,\right\rangle} 
\newcommand{\ket}[1]{|#1\rangle} 
\newcommand{\bra}[1]{\langle#1|} 

\newcommand{\vR}{\boldsymbol{R}}

\newcommand{\vecr}{\boldsymbol{r}}
\newcommand{\ve}{\boldsymbol{e}}
\newcommand{\vq}{\boldsymbol{q}}
\newcommand{\vk}{\boldsymbol{k}}

\begin{document}
\title{
Evaluating First-Principles Electron-Phonon Couplings: Consistency Across Methods and Implementations.
}

\author{Konrad Merkel\textsuperscript{1}}
\author{Maximilian F.X. Dorfner\textsuperscript{1}}
\author{Manuel Engel\textsuperscript{2}}
\author{Georg Kresse\textsuperscript{3}}
\author{Frank Ortmann\textsuperscript{1}}\email{frank.ortmann@tum.de}

\affiliation{\textsuperscript{1}Department of Chemistry, TUM School of Natural Sciences, and Atomistic Modeling Center, 
	Munich Data Science Institute, Technical University of Munich, Germany}

\affiliation{\textsuperscript{2}VASP Software GmbH, Vienna, Austria}
\affiliation{\textsuperscript{3}	
	University of Vienna, Faculty of Physics and Center for Computational Materials Science, Vienna, Austria}

\begin{abstract}
Electron-phonon coupling (EPC) is fundamental for understanding the behavior of molecules and crystals, influencing phenomena such as charge transport, energy transfer, phase transitions, and polaron formation. Accurate computational methods to calculate EPCs from first principles are essential, but their complexity has resulted in a variety of computational strategies, raising concerns about their mutual consistency.
In this study, we provide a systematic benchmark of methods for EPC calculation by comparing two fundamentally different {\it ab initio} methodologies. We investigate Gaussian-type orbital methods based on the \textsc{CP2K} code and plane-wave-based projector-augmented-wave (PAW) methods combined with maximally localized Wannier functions, as implemented in \textsc{VASP} and \textsc{wannier90}. In addition, we further distinguish between the derivative--of--Hamiltonian ($dH$) and  derivative--of--states ($d\psi$) approaches for obtaining EPC parameters. The comparison is conducted on a representative set of organic molecules, including pyrazine, pyridine, bithiophene, and quarterthiophene, varying significantly in size and flexibility.
We find excellent agreement across implementations and basis sets when employing the same computational approach ($dH$ or $d\psi$), demonstrating robust consistency between the numerical schemes. However, noticeable deviations occur when comparing the $dH$ and $d\psi$ approaches within each code and for specific cases discussed in detail.
Our findings emphasize the reliability of EPC computations using the $dH$ method and caution against potential pitfalls associated with the $d\psi$ approach, providing guidance for future EPC calculations and model parameterizations.
\end{abstract}

\maketitle

\section{Introduction}
Understanding the behavior of molecules and crystals requires consideration of both electronic and vibrational degrees of freedom, as well as their mutual interactions. Vibrational motion causes dynamical changes in the relative positions of atoms, thereby affecting the electronic structure. Electron-phonon interaction is central to a wide range of phenomena, including charge, energy or heat transport, 
luminescence, and relaxation processes.\cite{Bredas_2002,Coropceanu_2007,Troisi2006A,Fratini_2009,OrtmannPRB2009,Hutsch_2022,Hutsch_2024,Bredas2004,Hestand2018,Spies2025,Ziman_2001,Liao_2015,Zhou2020} It is also crucial for iconic effects such as phase transitions,\cite{Budai_2014,Luo_2022,BiancoCalandra_2015PRB} polaron formation, and superconductivity.\cite{Bardeen_1957A,Bardeen_1957B,Nambu_1960,Eliashberg_1960} For instance, coupling to high-frequency phonon modes can lead to the formation of polarons—quasiparticles consisting of a charge carrier dressed by a polarization cloud, which strongly influences charge transport. \cite{Holstein_2, Mahan, Hannewald2004_2, OrtmannPRB2009} 
Coupling to low-frequency phonon modes, on the other hand, can introduce vibrational disorder, charge scattering, and hopping-like transport. \cite{Troisi2006B, Troisi2011,Illig_reducing_2016} In many organic materials, both mechanisms coexist and govern the overall charge transport behavior. \cite{Merkel2022,Panhans2023}

To describe electronic and vibrational dynamics simultaneously,  Holstein-Peierls type of models have become a cornerstone across various scientific fields due to their flexibility and success in capturing essential material properties. \cite{Holstein_1, Mahan,KoppelDomcke-1984} While standard density-functional theory (DFT) codes are widely used to approach electronic structures and phonon spectra, computing electron-phonon couplings (EPCs) remains challenging,\cite{Giustino_2017} as it involves the simultaneous consideration of electronic and vibrational degrees of freedom. Several computational strategies have been developed, employing vastly different theoretical formalisms and basis sets. 
\cite{DFPTBaroni2001,Piscanec2004,NoffsingerEPW2010,Ponce2016,Chaput2019,Engel2020,Engel2022,Dorfner2025}.
Despite the growing number of techniques developed for this purpose, comprehensive benchmarks across different methodologies are still missing. Yet, such comparisons are essential, because model parametrization is frequently the first step toward large-scale simulations and, if different methods yielded inconsistent results, subsequent scientific findings would be questionable. Given the foundational role of electron-phonon parameters and the necessity for consistency across different computational schemes, establishing general agreement on these parameters is imperative.

In this work, we address this critical gap by systematically comparing two fundamentally different approaches for calculating EPCs. We investigate methodologies based on Gaussian basis sets using the \textsc{CP2K} code on the one hand and those using the projector-augmented wave (PAW)\cite{Blochl1994PRB,Joubert1999PRB} formalism combined with maximally localized Wannier functions, as implemented in \textsc{VASP}\cite{VASP1,VASP2,VASP3,VASP4} and \textsc{wannier90},\cite{Marzari1997,Marzari2012} on the other. Furthermore, by examining two different approaches for each method, we are able to distinguish between methodological and approach-specific effects, and assess their performance across a set of molecules of varying sizes and flexibility.

\section{Methods and Materials}
\subsection{Hamiltonian and Interactions}
The Holstein-Peierls Hamiltonian consists of three parts,
\begin{align}
\hat{H} =& \hat{H}_\text{el} + \hat{H}_\text{ph} + \hat{H}_\text{el-ph},
\end{align}
where $\hat{H}_\text{el}$ contains the electronic structure, $\hat{H}_\text{ph}$ the phonons and $\hat{H}_\text{el-ph}$ the electron-phonon interaction. The model can be used for both molecules and crystals.
For convenience, molecules can also be studied within a slab approach using an artificial crystal with wide vacuum regions between them to avoid interactions and restricting the crystal momentum to the $\Gamma$-point of the Brillouin zone.\cite{Bechstedt2015}

Vibrational degrees of freedom are included in terms of harmonic oscillators,
\begin{align}
\hat{H}_\text{ph}=& \sum_{\nu \vq} \hbar \omega_{\nu \vq} \left(\hat{b}^\dagger_{\nu \vq} \hat{b}_{\nu \vq} + \frac{1}{2}\right),
\end{align}
where $\hat{b}^{(\dagger)}_{\nu \vq}$ are (creation) annihiliation operators for a phonon mode $\nu$ with wave vector $\vq$ and frequency $\omega_{\nu \vq}$. Higher order terms like non-harmonic contributions are a possible extension of the Holstein-Peierls model but not in the scope of the present work. 

The electronic structure can be described in terms of Kohn–Sham quasiparticle eigenstates evaluated at fixed nuclear equilibrium positions, corresponding to molecular orbitals or Bloch states,
\begin{align}
\hat{H}_\text{el} = \sum_{n\vk} \epsilon_{n\vk} \hat{c}^\dagger_{n\vk} \hat{c}_{n\vk},
\label{eq:H_el_bloch}
\end{align}
where $\epsilon_{n\vk}$ are the energies of band $n$ and crystal momentum $\hbar\vk$. $\hat{c}^{(\dagger)}_{n\vk}$ are the associated (creation) annihilation operators for the Bloch states.
The electronic parameters depend on the nuclear positions $\vecr_{\kappa}$, where $\kappa$ is the atom index. 
By expressing the atomic displacement out of equilibrium $\Delta \vecr_{\kappa}$ in terms of phonon creation and annihilation operators we obtain the EPC part of the Hamiltonian as \cite{Holstein_1, Mahan}
\begin{align}
\hat{H}_\text{el-ph}=& \sum_{nm\vk} \sum_{\nu \vq} \hbar \omega_{\nu \vq} g_{mn\vk,\nu \vq} \left(\hat{b}^\dagger_{\nu \vq} + \hat{b}_{\nu -\vq}\right)\hat{c}^\dagger_{n\vk} \hat{c}_{m\vk+\vq}
\label{eq:el-ph-coupling-hamiltonian}
\end{align}
where $g_{mn\vk,\nu \vq}$ are dimensionless EPC parameters in the Bloch basis. They are defined as
\begin{align}
g_{mn\vk,\nu \vq}&= \sqrt{\frac{1}{2N_\Omega\hbar\omega_{\nu\vq}^3}} \bra{\psi_{m,\vk+\vq}}(\partial_{\nu,\vq} H) \ket{\psi_{n,\vk}} \label{eq:g_dH_bloch}
\end{align}
where $\partial_{\nu,\vq}$ denotes the derivative with respect to the coordinate of phonon mode $\nu$ and phonon wavevector $\vq$. 
We note that we use dimensionless coupling constants throughout this work.
$\hat{H}_\text{el-ph}$ can also be expressed in terms of Cartesian displacements of atoms 
\begin{align}
\partial_{\nu,\vq} = \sum_{\kappa\alpha} \frac{1}{\sqrt{M_\kappa}}\ve_{\kappa\alpha,\nu \vq} e^{i\vq \vR_{\kappa}} \frac{\partial}{\partial \vecr_{\kappa\alpha}},  
\label{eq:transformcoordinate}
\end{align}
where $\kappa$ labels the atom, $\alpha$ the Cartesian coordinate, $\ve_{\kappa\alpha,\nu \vq}$ the phonon mode pattern and $\vR_{\kappa}$ is the unit cell vector associated to atom $\kappa$.
Eq.~\eqref{eq:g_dH_bloch} already suggests a numerical evaluation, in which all atoms are displaced according to the phonon mode pattern using finite displacement and  $\partial_{\nu,\vq} H$ is calculated directly. Alternatively, one can displace each atom along its Cartesian directions and take a linear combination with respect to the phonon mode pattern afterwards.

We can reformulate Eq.~\eqref{eq:g_dH_bloch} and get an alternative formulation in terms of derivatives of the eigenstates and eigenenergies,
\begin{align} \label{eq:g_dPsi}
g_{mn\vk,\nu \vq}&= \sqrt{\frac{1}{2N_\Omega\hbar\omega_{\nu\vq}^3}} \left[\delta_{mn} \delta_{\vk, \vk+\vq} \partial_{\nu,\vq} \epsilon_{n\vk} + \left( \epsilon_{n\vk}- \epsilon_{m,\vk+\vq} \right) \bra{\psi_{m,\vk+\vq}} \left( \partial_{\nu,\vq} \ket{\psi_{n,\vk}} \right) \right].
\end{align}
A detailed derivation in shown in the appendix. In this form we can distinguish two contributions on the right hand side, i.e. the derivative of the eigenenergies in the first term and the derivative of the eigenstates in the second term. In the case of \emph{intrastate} coupling (where $m=n$, $\vq=0$) only the first term contributes, while, otherwise, only the second term contributes.

Eqs.~\eqref{eq:g_dH_bloch} and~\eqref{eq:g_dPsi} represent two independent approaches how the EPC parameters can be calculated numerically. For a precise distinction we want to call them  \emph{$dH$ approach} and  \emph{$d\psi$ approach} respectively. Although, the underlying formulas are mathematically equivalent, both approaches are different from an implementation point of view. In particular, the latter is quite appealing since energies and states are generally easier to access. That is, calculating energies and states for the relaxed and displaced geometries and employing Eq.~\eqref{eq:g_dPsi} as a post-processing step might be easier than calculating the derivative of the Hamiltonian matrix in Eq.~\eqref{eq:g_dH_bloch}.
On the other hand, difficulties may occur for the $d\psi$ approach when degenerate or nearly degenerate states are involved in the computations, for which the atomic displacement could lead to a mixing of different states that is not always obvious. Ignoring this could easily lead to wrong derivatives and therefore wrong coupling constant for such states. In such situations the $dH$ approach should be more robust since the expectation value is only calculated for electronic eigenstates of the equilibrium structure.

\subsection{Wannier basis and tight-binding-like models}

As an alternative to the representation in the form of quasiparticle eigenstates (Bloch states) of the electronic Hamiltonian discussed above, a localized real-space basis in terms of maximally localized Wannier functions (MLWFs) might be more suitable for specific applications. MLWFs $\ket{w_{i\vR}}$ are defined based on a transformation of Bloch states $\ket{\psi_{n\vk}}$, specifically,
\begin{align} \label{eq:wannier_trafo}
\ket{w_{j\vR}} &= \frac{1}{\sqrt{N_\Omega}} \sum_{n\vk} e^{-i\vk \vR} U_{jn,\vk} \ket{\psi_{n\vk}},
\end{align}
where $\vR$ is a unit cell vector and $U_{jn,\vk}$ is a unitary that fixes the $k$-depended gauge phase of Bloch states such that the obtained Wannier functions are maximally localized. It can be obtained using established techniques. \cite{Marzari1997,Souza2001, Marzari2012} The choice of representation -- Bloch or Wannier basis -- may depend on the system and the specific purpose or observable of interest. Due to their strong localization, MLWF usually lead to very sparse representations of the Hamiltonian, which is beneficial for large scale computations.

The electronic part of the Holstein-Peierls Hamiltonian Eq.~\eqref{eq:H_el_bloch} can be written in terms of MLWF as
\begin{align}
\hat{H}_\text{el} = \sum_{ij\vR\vR'} \epsilon^{\text{W}}_{ij,\vR'-\vR} \hat{a}^\dagger_{i\vR} \hat{a}_{j\vR'},
\end{align}
where the matrix elements $\epsilon^{\text{W}}_{ij,\vR'-\vR}$ denote the transfer integrals and onsite energies of the equivalent tight-binding-like model. $\hat{a}^{\dagger}_{i\vR}$ and $\hat{a}^{}_{j\vR'}$ are the creation and annihilation operators for Wannier orbitals, respectively, with orbital index $i$ and lattice vector $\vR$.
Similarly, the EPC term in the Hamiltonian is given by
\begin{align}
\hat{H}_\text{el-ph}=& \sum_{ij\vR\vR'} \sum_{\nu \vq} \hbar \omega_{\nu \vq} g^{\text{W}}_{ij\vR\vR',\nu \vq} \left(\hat{b}^\dagger_{\nu \vq} + \hat{b}_{\nu, -\vq}\right)\hat{a}^\dagger_{i\vR} \hat{a}_{j\vR'},
\label{eq:el-ph-coupling-hamiltonian-Wannier}
\end{align}
where $g^{\text{W}}_{ij\vR\vR',\nu \vq}$ are EPC constants in the Wannier basis. They are defined in close analogy to Eq.~\eqref{eq:g_dH_bloch} as
\begin{align}
g^{\text{W}}_{ij\vR\vR',\nu \vq} &= \sqrt{\frac{1}{2N_\Omega\hbar\omega_{\nu\vq}^3}} \bra{w_{i\vR}}(\partial_{\nu,\vq} H) \ket{w_{j\vR'}} \label{eq:g_dH_wannier}
\end{align}

Using the Wannier transformation Eq.~\eqref{eq:wannier_trafo} we can easily relate them to the EPC in Bloch representation by
\begin{align} \label{eq:g_trafo_wannier_bloch}
g_{mn\vk,\nu \vq} = \sum_{i\vR}\sum_{j\vR'} e^{-i(\vk+\vq)\vR}
U_{mi,\vk+\vq} g^{\text{W}}_{ij\vR\vR',\nu \vq} U^\dagger_{jn,\vk} e^{i\vk \vR'}.
\end{align}
The inverse transform is in principle also possible. However, in a practical calculation one needs to assure that the $k$-dependent gauge phase of all Bloch functions (which was fixed during the wannierization process) remains the same in all calculations. This is usually not assured.

\subsection{Materials}
\begin{figure}
	\centering
	\includegraphics[width=0.4\textwidth]{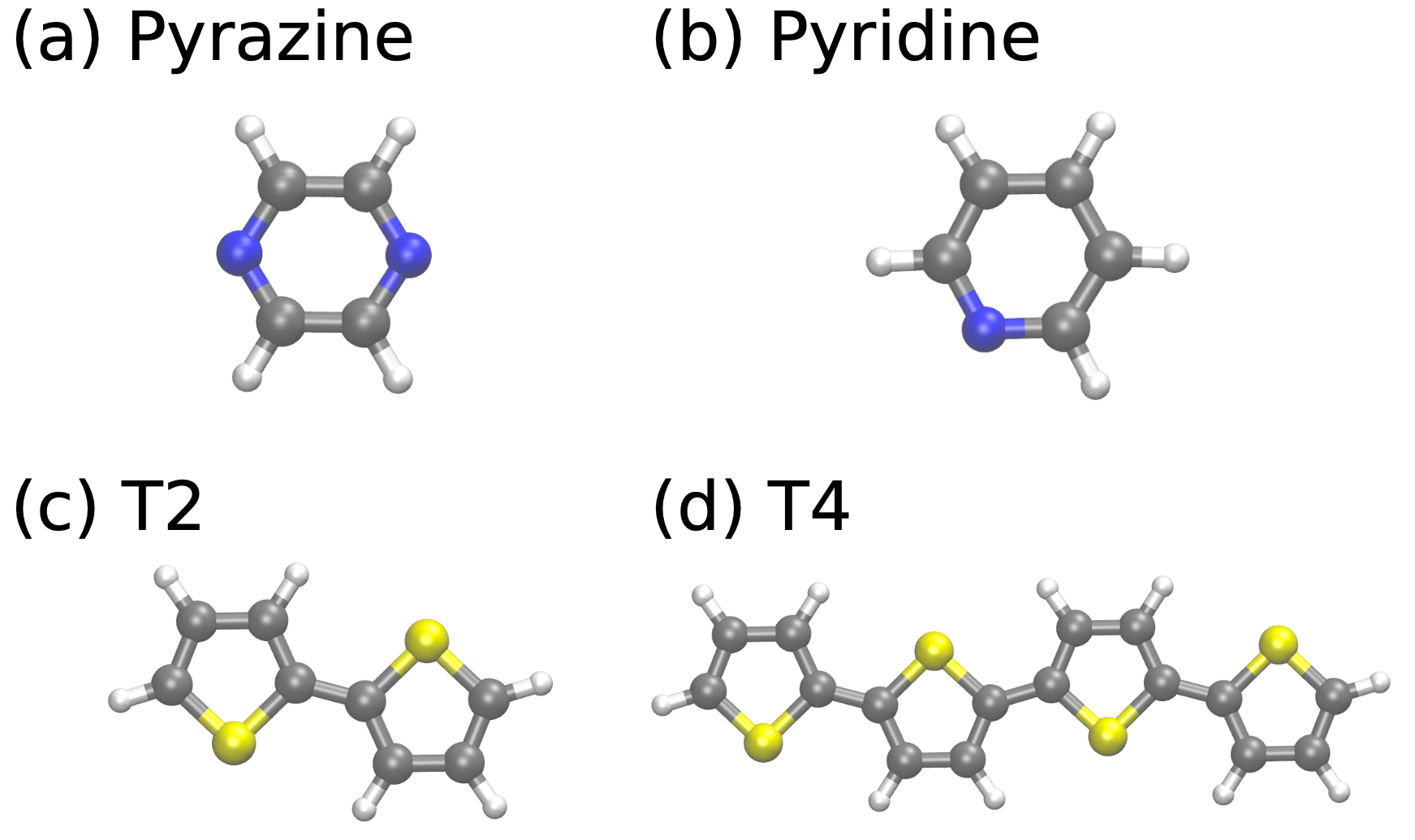}
	\caption{Molecules used for the comparison of EPCs, with atoms colored by element: gray for carbon, white for hydrogen, blue for nitrogen, and yellow for sulfur.}
	\label{fig:materials}
\end{figure}

For our investigation, we study molecules in slab geometry. We choose four molecules with heteroatoms of varying size, namely pyrazine, pyridine, bithiophene (T2) and quarterthiophene (T4), which are shown in Fig.~\ref{fig:materials}. This set therefore includes small rigid molecules with only about 10 atoms and larger structures with up to 30 atoms. The largest one T4 has three dihedral angles between the thiophene fragments and exhibits the strongest flexibility.

\section{Computational Details}

\begin{table}
	\centering
	\caption{The numerical evaluation of the EPC parameters is quite different between \textsc{CP2K} and \textsc{VASP}. The table compares the most important aspects of the \emph{ab initio} methodologies.}
	\label{tab:diff}
	\begin{tabular}{ccc}
		\hline
		\hline
		& \textbf{\textsc{CP2K}} &  \textbf{\textsc{VASP}} \\
		\hline
		DFT-functional	& PBE & PBE \\
		Basis set	& Gaussian & PAW + Wannier functions \\
		Evaluation method for DFT & GPW & PAW \\
		Pseudo-potentials & GTH & GW flavour \\
		EPC parameter	& in Kohn-Sham basis & in Wannier basis \\
		Displacement of atoms &  along phonon mode pattern  & along Cartesian directions \\
		Periodic boundary conditions	& no periodicity & large vacuum between molecules, only $\Gamma$-point\\
		\hline
		\hline
	\end{tabular}
\end{table}

All {\it ab-initio} calculations are performed using density functional theory (DFT) \cite{Kohn_1965} and the PBE functional.\cite{Perdew_1996_A} A comparison of other functionals can be found elsewhere.\cite{Faber2011PRB,Dorfner2025} 
For the DFT calculations, we employ the \textsc{CP2K} and \textsc{VASP} codes in this study. For \textsc{VASP}, both the $dH$ and $d\psi$ approaches for calculating the coupling constants are supported directly. For \textsc{CP2K}, these approaches are carried out using a custom-made package, interfaced with \textsc{CP2K}. The latter is used for the single-point calculations, as used previously in Ref. \cite{Dorfner2025}. Currently this is not a built-in feature of \textsc{CP2K}. However, for simplicity and clarity, we still call this approach CP2K in the following.
Having these separate numerical implementations prepared, enables a systematic investigation of differences arising from the two approaches themselves, as well as from the use of different {\it ab-initio} methodologies within a given approach. For convenience, we summarize the most important differences in Tab. \ref{tab:diff}.

The calculation of phonon mode patterns and phonon frequencies is done in a first step in each case. For this, we use a finite displacement of 0.03~Bohr (=~\SI{0.01588}{\angstrom}) to obtain the dynamical matrix by a central difference approach to the forces. The matrix is then diagonalized numerically.

Now we come to the calculation of the EPCs.
In \textsc{VASP}, the $dH$ approach is implemented as described in Ref.~\cite{Engel2020}. For the comparison of EPC matrix elements in this work, we examine the all-electron (AE) matrix elements  described in Ref.~\cite{Chaput2019, Engel2022}. The calculation is based on MLWF from all valence bands, which are calculated using \textsc{VASP} and \textsc{wannier90}. The EPC couplings are then calculated directly in the Wannier basis with respect to Cartesian displacements, i.e. we obtain
\begin{align}
	\overline{g}^{\text{W}}_{ij\vR\vR',\kappa\alpha}&= \bra{w_{i\vR}}(\partial_{\kappa\alpha} H) \ket{w_{j\vR'}},
\end{align}
where $\kappa$ is the atom index and $\alpha$ the Cartesian component. This can be transformed into coupling constants with respect to phonon mode index and momentum by using the mode pattern and applying Eq.~\eqref{eq:transformcoordinate}.

The $d\psi$ approach is implemented in \textsc{VASP} based on Eq.~\eqref{eq:g_dPsi}, which is further expressed in the language of the PAW method.
The overlap matrix \(\braket{\psi_{m,\vk+\vq}}{\partial_{\nu,\vq} \psi_{n,\vk}} \) is evaluated in a supercell using only the $\Gamma$-point.
A first implementation approaching this overlap was discussed in Ref.~\cite{Turiansky2011}, where the authors computed the quantity
$\bra{\tilde{\psi}_{m}}
	\tilde{S}
	\ket{\tilde{\psi}_{n}^{\nu, \vq}}$,
with \(\bra{\tilde{\psi}_{m}}\) 
the pseudo (PS) orbitals of the undisplaced geometry, \(\ket{\tilde{\psi}_{n}^{\nu, \vq}}\) the perturbed or displaced PS orbital, and \(\tilde{S} \) the PAW overlap operator.
The derivative is readily computed by varying the amplitude of the mode displacement and fitting the linear dependence.
This PAW formulation is known as the so-called PS approach~\cite{Engel2022}, and should be clearly distinguished from later approaches as its result is not fully equivalent to our target quantity \(\braket{\psi_{m,\vk+\vq}}{\partial_{\nu,\vq} \psi_{n,\vk}} \).
Since our goal is to compare electron-phonon coupling constants, we use an AE formulation following the ideas in Refs.~\cite{Chaput2019, Engel2022}.
The PAW AE equivalent of the inner product in Eq.~\eqref{eq:g_dPsi} is computed as
\begin{align}
	\sum_{aij}
	\braket{\tilde{\psi}_{m}}{\tilde{p}_{ai}}
	\left(
		\braket{\phi_{ai}}{\phi_{aj}^{\nu, \vq}} -
		\braket{\tilde{\phi}_{ai}}{\tilde{\phi}_{aj}^{\nu, \vq}}
	\right)
	\braket{\tilde{p}_{aj}^{\nu, \vq}}{\tilde{\psi}_{n}^{\nu, \vq}}
	,
\end{align}
where \(\ket{\phi_{ai}} \) and \(\ket{\tilde{\phi}_{ai}} \) are the AE and PS partial waves, respectively, and \(\ket{\tilde{p}_{ai}} \) are the PAW projector functions.
The index \(a \) is an atom index while the indices \(i \) and \(j \) are the PAW channels, i.e., compound indices that refer to the angular momentum and magnetic quantum numbers in the PAW datasets. 
As before, the superscript \((\nu, \vq)\) indicate the perturbed states.
We note that the difference between the AE and PS coupling strengths only exists in \((m \neq n) \) off-diagonal matrix elements, which is why it suffices to consider only the inner product in Eq.~\eqref{eq:g_dPsi} for this discussion.
The perturbed projectors and PS orbitals are computed directly via a self-consistent VASP calculation in the displaced \((\nu, \vq)\) geometry.
The inner products involving partial waves, however, are computed from the gradient of the partial waves multiplied by the displacement pattern.
To give an idea, if we denote the displacement vector of atom \(a \) in the perturbed structure as \(\boldsymbol{\Delta}_{a \nu, \vq} \), then
\begin{align}
	\ket{\phi_{aj}^{\nu, \vq}}
	=
	\ket{\phi_{ai}}
	+
	\boldsymbol{\Delta}_{a \nu, \vq}
	\cdot
	\ket{\boldsymbol{\nabla} \phi_{ai}}
	.
\end{align}

Within \textsc{CP2K}, the calculation of the coupling constants, in either approach, is done by displacing the atoms according to the phonon mode patterns and calculating the resulting change in energy and electronic states, as done previously in Ref.~\cite{Dorfner2025}. There, the EPC constants are then directly obtained in the Kohn-Sham basis. Any appearing derivative along the normal mode coordinate is approximated by a central difference of the form 
\begin{equation}
		\partial_{\nu,\mathbf{0}}X\approx\frac{X^+-X^-}{2\delta \sqrt{M_{\nu,\mathbf{0}}}},
\end{equation}
where $X^{\pm}$ stands for the quantity $X$ in the system that has been displaced by $\pm\delta$ along the normal mode direction  $\mathbf{e}_{\nu,\mathbf{0}}$. This dimensionless quantity is obtained from the eigenvector of the mass-weighted Hessian $\xi_{\nu,\mathbf{0}}$, according to 
\begin{equation}
		\mathbf{e}_{\nu,\mathbf{0}}=\xi_{\nu,\mathbf{0}}\sqrt{M_{\nu,\mathbf{0}}},
\end{equation}
where $M_{\nu,0}=\vert\vert\xi_{\nu,\mathbf{0}}\vert\vert^{-2}$ is the corresponding reduced mass of the mode. 

In the $dH$ approach, $X$ stands for the Kohn-Sham Hamiltonian. After evaluation of this discretized derivative, we compute the matrix elements of this operator with the Kohn-Sham eigenstates $\ket{\psi_{m,\mathbf{0}}}$ to obtain the coupling constants. In contrast, in the $d\psi$ approach, $X$ denotes either the respective Kohn-Sham eigenvalue (for $m=n$) or the respective eigenstate of the Kohn-Sham Hamiltonian (for $m\neq n$). While in the case $m=n$ the coupling constants can be obtained directly from this discretized derivative, the case $m\neq n$ requires the computation of the overlap of the discretized derivative with eigenstates at the non-deflected geometry.

Since the atom-centered Gaussian-type orbital (GTO) basis functions $\{\ket{\phi_{x,\pm}}\}$ are varied slightly upon displacement, their associated quantities cannot be directly subtracted, as they refer to different bases. Instead, we first transform the displaced vectors or matrices back to the equilibrium (non-deflected) basis $\{\ket{\phi_{x}}\}$ using the analytic overlap matrix
\begin{align}
		T_{x,y}^{\pm}=\langle\phi_x\mid\phi_{y,\pm}\rangle.
\end{align} 
This ensures consistency when evaluating the finite  difference approximations.

When computing derivatives of the Kohn–Sham eigenstates, an additional complication can arise from level crossings. Specifically, as the system is slightly displaced, two eigenstates may hybridize or exchange their energetic ordering.  This can lead to ambiguity in identifying which perturbed (displaced) state corresponds to which original (unperturbed) state. To resolve this, we match the respective states by a maximum overlap procedure. For a detailed description of this procedure, we refer to Ref. \cite{Dorfner2025}.

\section{Results}

We start our comparison with the electronic structure and, subsequently, the phonon modes.
For the electronic structure we observe a very good agreement of the electron energies, where the largest deviation is found for the lowest--energy valence states with a maximal differences of \SI{45}{\milli\eV} (0.5\%) between \textsc{VASP} and \textsc{CP2K}. Next, the obtained Wannier functions reproduce the electronic structure perfectly with the largest deviation between the Wannier representation and the original Kohn-Sham eigenvalues from \textsc{VASP} of \SI{5e-9}{\eV}.
Using the Wannier functions, we can also compare the Kohn-Sham states on a real-space grid, which are obtained from a diagonalization of the Wannier-Hamiltonian. Comparing with the Kohn-Sham states from \textsc{CP2K} yields very good agreement even for states with small energy differences. Overall, we did not encounter any problems due to degeneracies or near-degeneracies.

For the phonon modes, both methodologies reproduce the same modes in terms of their frequencies and mode patterns. Deviations in energies are only found for low-frequency modes, which may be expected to some extend. Since these residual deviations between the phonons, even if small, could lead to deviations in the EPC constants, we avoid this unnecessary error propagation and continue by using only the mode patterns and frequencies from  \textsc{CP2K}. This means that we calculate all the couplings with exactly the same phonon energies and mode patterns to make the results fully comparable.

\begin{figure}
	\centering
\includegraphics[width=0.8\textwidth]{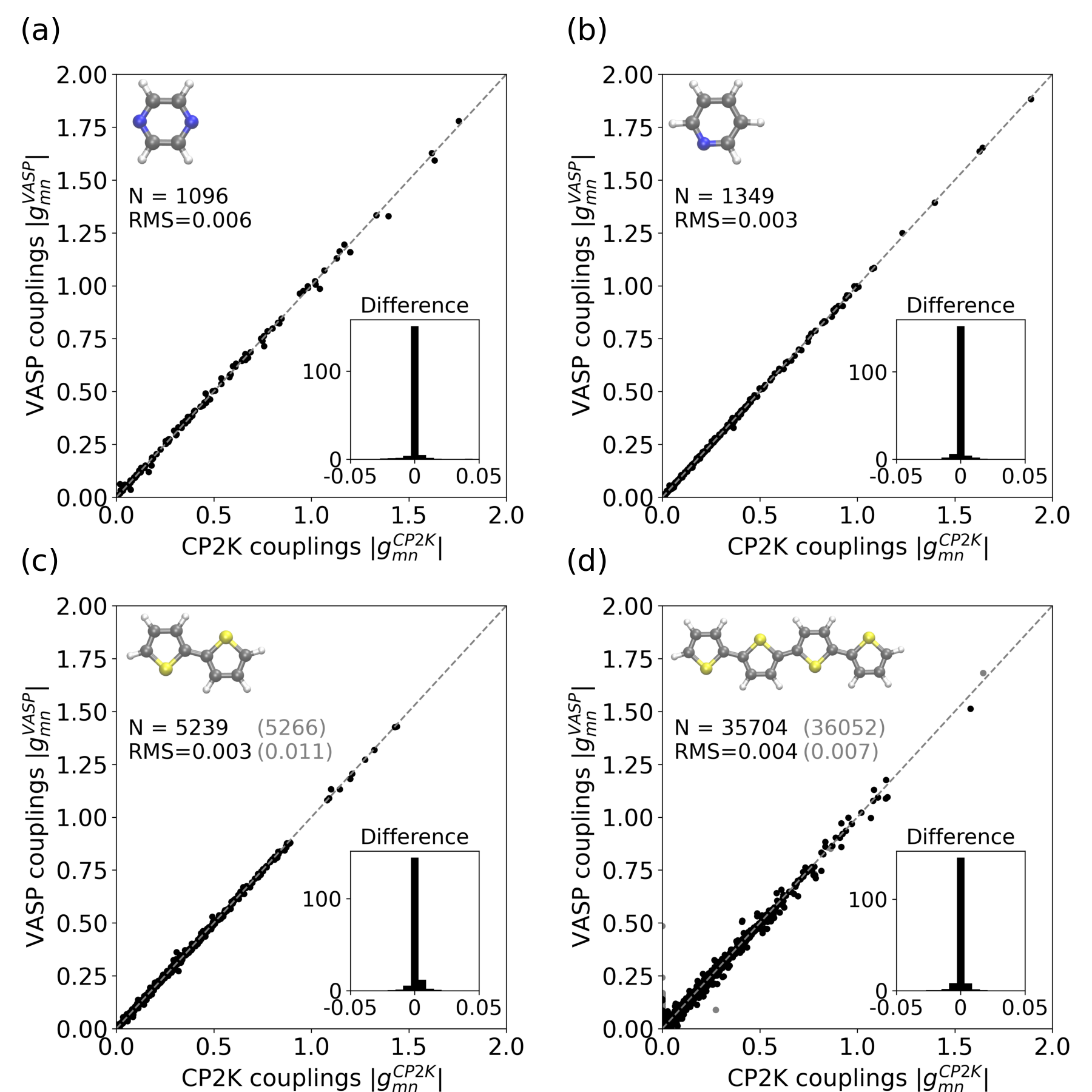}
	\caption{Comparison of EPCs calculated using \textsc{VASP} and \textsc{CP2K} for various molecules (see insets). Both codes employ the derivative of the Hamiltonian (cf. Eq.~\eqref{eq:g_dH_bloch}), referred to as $dH$ approach in the main text, but differ significantly in terms of wavefunction basis and implementation specifics (cf. Tab.~\ref{tab:diff}). 	
		In (c) and (d), the black points include all vibrations above $\SI{74}{\centi\meter^{-1}}$ with corresponding RMS (root mean square) deviation, while grey points belong to modes with wavenumbers $<\SI{74}{\centi\meter^{-1}}$. The RMS value for the entire set is indicated in parentheses.
		Note that the seemingly greater scatter of black points for the 4T molecule in the main panel of (d) is rather due to the much larger number of data points, while the RMS values are comparable.}
	\label{fig:comparison_dH}
\end{figure}

We continue with the comparison of the EPCs in the Kohn-Sham basis. In materials simulations that consider EPC, like charge transport or optical absorption/reflection calculations, the physically relevant excitations usually occur around the Fermi level, i.e. for the top valence or bottom conduction bands and possibly their neighbor bands. Therefore, it would be sufficient to obtain the EPC parameters only for these relevant bands. In general it depends on the specific simulations, the excitation energy and the material, how many bands are actually needed.
In this paper, we do not want to restrict our comparison to only a few states but rather use all valence states. This provides us with enough variety of electronic states for a meaningful discussion. However, we want to omit EPCs that are far beyond the Holstein-Peierls model and would require higher order corrections. For instance, couplings between highest and lowest valence state with electronic energy differences of $\approx\SI{16}{\eV}$, involving phonons whose energies are orders of magnitude smaller, are not in the focus here.
We have therefore decided to compare only those couplings where the differences between phonon energy and electronic energy differences are not too far apart. As a practical criterion, we require that $\Delta E_{\text{el}} < 10E_\text{ph}$, meaning that up to 10 phonons might be considered to enable a transition between electronic states. This is a very generous criterion as it does not imply too large restrictions. Note that even a looser condition of $\Delta E_{\text{el}} < 100 E_\text{ph}$ would yield similar results. 


\subsection{EPCs using the $dH$ approach}

To compare the EPCs, we first calculate them using the $dH$ approach, i.e. we obtain the couplings by evaluating Eq.~\eqref{eq:g_dH_bloch} and Eq.~\eqref{eq:g_dH_wannier} using \textsc{CP2K} or \textsc{VASP} respectively.
The results are shown in Fig.~\ref{fig:comparison_dH}. For all materials we can see that the results from both {\it ab initio} methodologies agree very well with each other, which is reflected in very low root mean square (RMS) error of 0.01 or even below. It is quite remarkable that such very different methods and codes, one using Gaussian basis functions, the other plane waves and Wannier functions, coincide so perfectly with each other for tens of thousands of coupling constants. The tiny deviations that can be observed are in the same order of magnitude as the relative variation in the electronic energies between \textsc{CP2K} and \textsc{VASP} that we calculated in the equilibrium geometries. They can be attributed to residual numerical errors that are inherent in each methodology and are typical for numerical calculations. 

We can also observe that larger molecules, such as T2 and T4, have somewhat larger numerical errors. This is reflected in slightly larger RMS values (in parentheses in Figs.~\ref{fig:comparison_dH} (c),(d)). These are mainly caused by a few low-frequency modes as evidenced by the black points and much lower RMS values when these modes are excluded.
 The possible occurrence of errors for larger molecules in case when atoms are displaced in Cartesian directions (instead of the phonon mode patterns) and individual components are summed (according to the phonon mode pattern) afterwards, has been discussed previously.\cite{OrdejonPRB2017} Still, the difference in the couplings between both {\it ab initio} methodologies is almost negligible and, despite the very different methods, basis sets and implementations, we find a highly satisfying agreement. 
It should be highlighted in this context that \textsc{VASP} calculates all EPC parameters in the Wannier basis, which are then transformed in the Kohn-Sham basis afterwards using Eq.~\eqref{eq:g_trafo_wannier_bloch}. Given the numerical simplicity of Eq.~\eqref{eq:g_trafo_wannier_bloch}, we can conclude that the coupling constants in the Wannier representation are of the same high quality. This freedom of representation is very advantageous since it gives researchers more flexibility in choosing the appropriate basis (Wannier or Kohn-Sham) for their specific needs. On the other hand taking the back-transformation from Kohn-Sham to Wannier couplings might not be straightforward because the $k$-dependent gauge phase of Bloch functions needs to be fixed as discussed above.

\subsection{EPCs using the $d\psi$ approach}

\begin{figure}
	\centering
\includegraphics[width=0.8\textwidth]{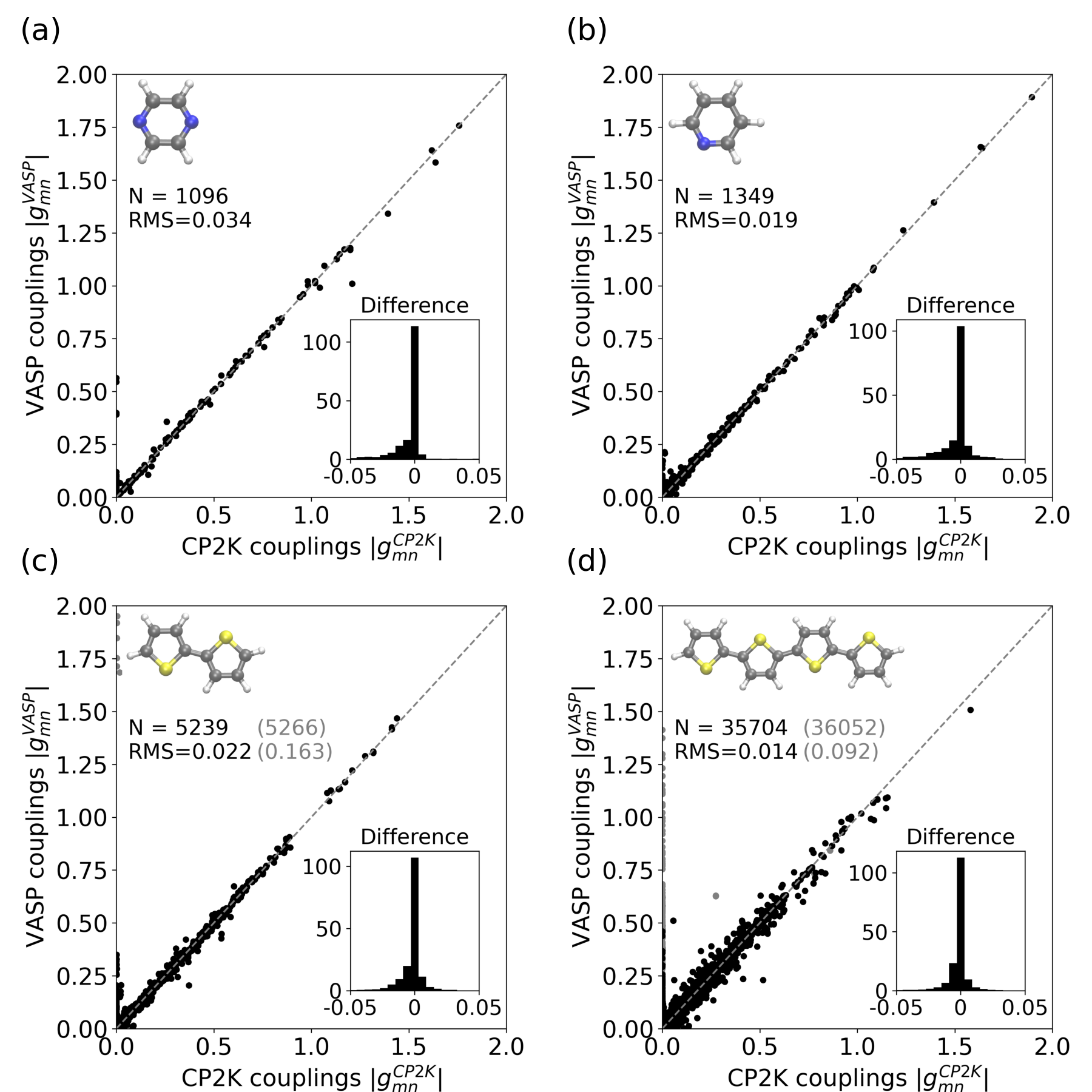}
	\caption{Comparison of EPCs between \textsc{VASP} and \textsc{CP2K} for various materials based on the derivative of states and eigen-energies (cf. Eq.~\eqref{eq:g_dPsi}), referred to as $d\psi$ approach in the main text. 	In (c) and (d), the black points include all vibrations above $\SI{74}{\centi\meter^{-1}}$ with corresponding RMS (root mean squared) deviation, while grey points belong to modes with wavenumbers $<\SI{74}{\centi\meter^{-1}}$. The RMS value for the entire set is indicated in parentheses. Note that the seemingly greater scatter of black points for the 4T molecule in the main panel of (d) is simply due to the much larger number of data points, while the RMS values are comparable.}
	\label{fig:comparison_dPsi}
\end{figure}

As an alternative to calculating the derivative of the Hamiltonian matrix, the $d\psi$ approach calculates the EPC from the derivative of the eigenstates and energies using Eq.~\eqref{eq:g_dPsi}. In this approach one only needs to calculate the derivative of the Kohn-Sham energies as well as the scalar product between states of the displaced and equilibrium geometries. This can be done directly in the PAW or Gaussian basis as a post-processing step without using Wannier functions. We therefore compare only the coupling constants of the Kohn-Sham states.
Unlike above cases, both implementations take the displacement along the phonon mode pattern directly and do not use Cartesian displacements.

Fig.~\ref{fig:comparison_dPsi} compares the resulting EPC constants in the $d\psi$ approach. As before, a very good agreement between the very different methodologies can be seen. The agreement is only slightly worse compared to the $dH$ approach and a somewhat larger scatter can be recognized, while the RMS errors between 0.01--0.03 are still very small when low-frequency modes ($<\SI{74}{\centi\meter^{-1}}$) for T2 and T4 are disregarded. 
For all four molecules we encounter a few (black) points close to the y-axis. A detailed analysis revealed that they are due to the choice of numerical differentiation methods, which are forward-difference for \textsc{VASP} and central-difference for \textsc{CP2K}. This can be fixed by using the reverse displacement for the \textsc{VASP} calculation as well to obtain the central difference in a post-processing step. By doing so, we find much better agreement. This procedure, however, also requires that phase shifts of independent simulations be aligned, which is not guaranteed in \textsc{VASP}. Similar to above the low-frequency modes in T2 and T4 (gray dots in Fig.~\ref{fig:comparison_dPsi}) determine the RMS values calculated for the entire set of modes, which indicates that these few modes deserve particular attention and careful treatment in any calculation of EPC parameters.
For instance, for the lowest-frequency mode in T2, the $d\psi$ approach incorrectly suggest EPC values close to 2 although they should be close to zero (upper left corner in Fig.~\ref{fig:comparison_dPsi} (c)), implying that using the $dH$ approach is crucial here.

\subsection{Comparison of $d\psi$ and $dH$ approaches}

\begin{figure}
	\centering
\includegraphics[width=\textwidth]{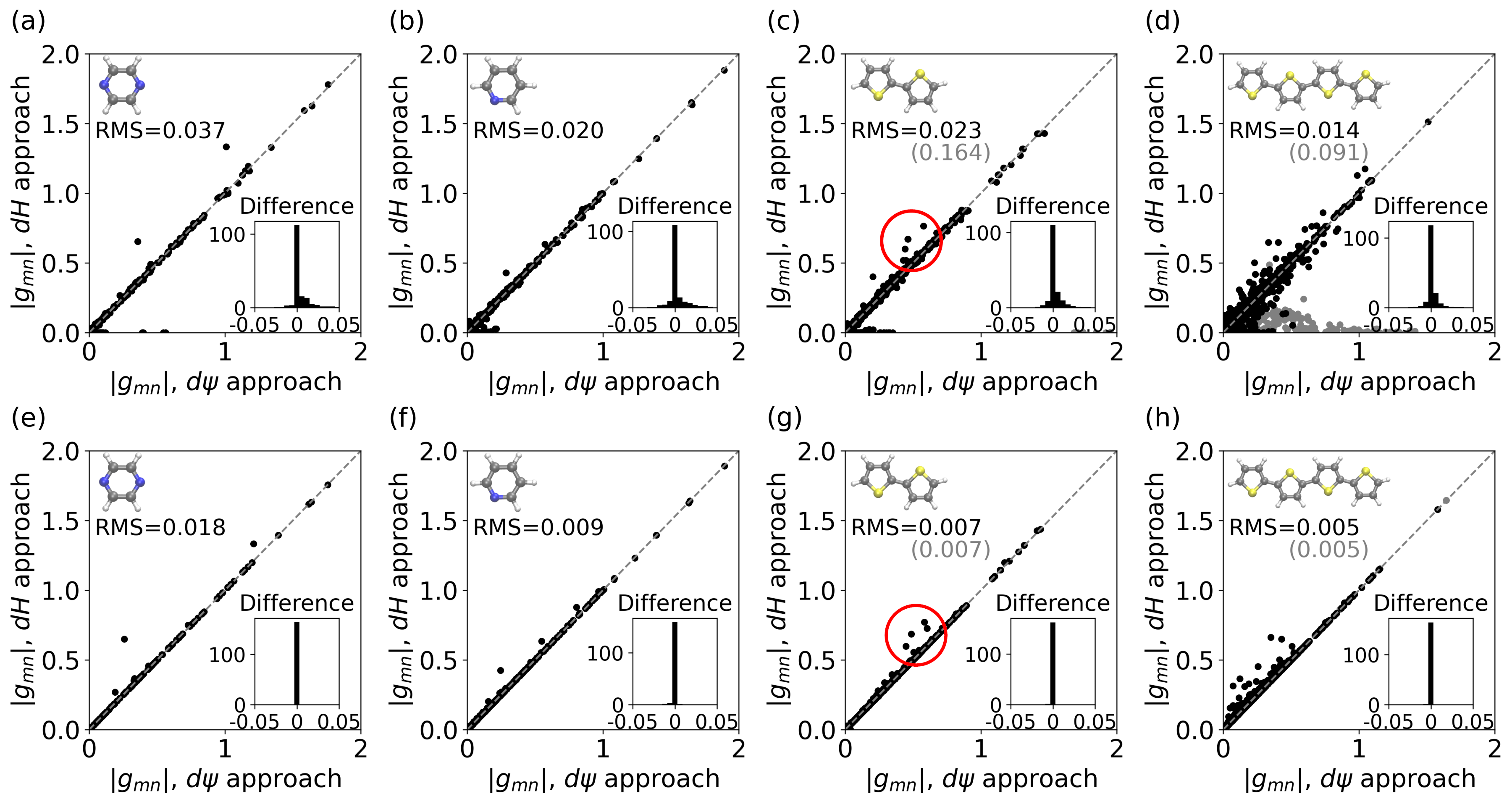}
	\caption{Comparison between $dH$ and $d\psi$ approaches within each code. Results are from \textsc{VASP} (a-d) and \textsc{CP2K} (e-h). Low-frequency modes for T2 and T4 ($<\SI{74}{\centi\meter^{-1}}$) are indicated as gray symbols. RMS values in parenthesis include all modes. The number $N$ of data points in each panel is the same as in corresponding previous figures.}
	\label{fig:comparison_methods}
\end{figure}

\begin{table}
	\centering
	\caption{Outliers for the $dH$--$d\psi$ comparison of EPC parameters for T2.}
	\label{tab:outliers}
	\begin{tabular}{ccccccccc}
		\hline
		\hline
		\textbf{electr. } &  \textbf{electr. } &  ~$E_2-E_1$~ &  ~\textbf{phonon mode}~ & \multicolumn{2}{c}{$dH$} &~& \multicolumn{2}{c}{$d\psi$ } \\
	\cline{5-6} \cline{8-9}
		\textbf{state 1} & \textbf{state 2} &&& $g^{\text{VASP}}$ & $g^{\text{CP2K}}$ &&  $g^{\text{VASP}}$ & $g^{\text{CP2K}}$ \\
		\hline
		13 & 14 & \SI{ 13.7 }{\milli\eV} & 29 (\SI{673.6}{\centi\meter^{-1}}) & 0.76 & 0.77 && 0.58 & 0.58 \\
		13 & 14 & \SI{ 13.7 }{\milli\eV} & 25 (\SI{796.7}{\centi\meter^{-1}}) & 0.60 & 0.60 && 0.44 & 0.45 \\
		17 & 18 & \SI{ 49.9 }{\milli\eV} & 19 (\SI{885.5}{\centi\meter^{-1}}) & 0.72 & 0.73 && 0.67 & 0.60 \\
		15 & 16 & \SI{ 64.1 }{\milli\eV} & 7 (\SI{1485.7}{\centi\meter^{-1}}) & 0.67 & 0.69 && 0.46 & 0.49 \\
		\hline\hline
	\end{tabular}
\end{table}

So far, the results are reassuring: the employed {\it ab initio} methodologies, at their convergence settings, yield consistent coupling constants -- with the $dH$ approach safely giving values that are essentially equal in both codes.
In addition, we aim to compare the results of the $dH$ and $d\psi$ approaches directly and plot the couplings in Fig.~\ref{fig:comparison_methods}. While the few low-frequency modes (gray points) have been already discussed above, we focus on the remaining large set of points.

Somewhat surprisingly, Fig.~\ref{fig:comparison_methods} exhibits some deviations between the approaches also for these modes. This occurs for \textsc{VASP} (Fig.~\ref{fig:comparison_methods}(a)-(d)) and \textsc{CP2K} (Fig.~\ref{fig:comparison_methods}(e)-(h)) independently. While clearly more than 99\% of the couplings are still in perfect agreement, we do see a few outliers, exemplarily encircled in panels (c) and (g). Interestingly, these outliers are similarly reproduced by \textsc{VASP} and \textsc{CP2K}, indicating that their origin is different from effects discussed above. This observation is also rather surprising since the mathematical expressions underlying both approaches are formally equivalent, but the implementations differ significantly.

This is best seen for the T2 molecule, which we select for a more detailed analysis. In Fig.~\ref{fig:comparison_methods} (c) and (g) we highlight four outliers at around $g=0.6$. In this case, the deviations between $dH$ and $d\psi$ values are not only of similar magnitude in both panels, they also correspond to the same coupling constants (i.e., identical state and mode indices for both methodologies), as listed in Tab.~\ref{tab:outliers}. The table further shows that each of these outliers corresponds to a coupling involving a distinct phonon mode, with mode energies ranging widely from  $\SI{673.6}{\centi\meter^{-1}}$ to $\SI{1485.65}{\centi\meter^{-1}}$, clearly belonging to higher-frequency modes rather than low-frequency ones. Furthermore, each outlier involves a different pair of electronic states (apart from the first two), with energy differences ranging from \SI{13.7}{\milli\eV} to \SI{64.1}{\milli\eV}. This indicates that the observed mismatches are not caused by a single electronic state or a single vibration but can arise from different combinations. Interestingly, all other couplings that correspond to these electronic states or these phonon modes show no deviations. That is, only the specific combinations of states with specific phonon modes cause problems.

Motivated by these observations, we sought to identify the origin of the outliers. We first tested different numerical parameters for each calculation and found that some outliers are slightly affected by the finite displacement amplitude. Changing this parameter changes the position of the outlier but no matter how we change this parameter it does not lead to better overall agreement. On the other hand, increasing the energy cutoff as an important convergence parameter for DFT implementations, does not lead to any changes in both codes. We also tried using smaller augmentation spheres for the PAW formalism in the \textsc{VASP} calculations and could not find any significant effect. Also a detailed analysis of the electronic states at equilibrium geometries has not revealed any significant differences.
The discrepancy only appears when the geometry is displaced along a specific phonon mode. In these cases, we observed a mixing of electronic states -- an effect that arises only in the $d\psi$ approach, since it is the only method that explicitly depends on the displaced eigenstates.
These are not used in the $dH$ approach, as explained above. 

We also find that the electronic mixing does not only occur between both involved states (states $m$ and $n$ for coupling $g_{mn,\nu}$). It might also happen that one of the involved states (say $m$) mixes with another state (say $m'$) that becomes close in energy upon phonon displacement. Interestingly, the \textsc{VASP} and \textsc{CP2K} methodologies, despite being different, exhibit the same mixing behavior, resulting in very similar deviations of the coupling constants, as demonstrated by the four outliers presented here. This mixing behaviour is thus a robust difference between both approaches.
The potential for state mixing is therefore a drawback of the $d\psi$ approach, which does not occur in the $dH$ approach. 
Furthermore, we found that in case of significant mixing the relation
$\left(\partial_{\nu,\vq} \bra{\psi_{m,\vk+\vq}}\right) \ket{\psi_{m,\vk+\vq}} = -\bra{\psi_{m,\vk+\vq}} \left( \partial_{\nu,\vq} \ket{\psi_{n,\vk}}\right)$ is violated numerically and the coupling constants are not symmetric with respect to the electronic states $g_{mn,\nu} \ne g_{nm,\nu}$, as they should be.

\section{Conclusion}

EPC constants can be calculated numerically either using the 
derivative--of--Hamiltonian ($dH$) or the  derivative--of--states ($d\psi$) approach. Both approaches are implemented in \textsc{VASP} and accessible by using \textsc{CP2K}, which are DFT codes based on very different methodologies in terms of basis sets, numerical implementation and calculation schemes.
Although the implementations are very different, the overwhelming majority of EPC constants are in excellent agreement as long as we use the same numerical approach (either $dH$ or $d\psi$). The lowest-frequency modes, however, can be more challenging but can be treated most accurately in the $dH$ approach as demonstrated in Fig.~\ref{fig:comparison_dH}. 
Comparing these two approaches with each other shows a few outliers that are consistently reproduced in both implementations. This is surprising since $d\psi$ and $dH$ approaches are mathematically equivalent, however, in practical calculations the $d\psi$ approach might suffer from mixing of electronic states upon displacements of the geometry.
Overall, we therefore would recommend to use the $dH$ approach if available. When using the $d\psi$ approach, one should verify the results critically and check if the coupling constants are symmetric with respect to electronic states, which may serve as a measure of quality. We expect that the implementation of the $dH$ approach would generally prove advantageous for any DFT code that allows the calculation of EPC parameters.

\section{Appendix}

\subsection{Derivation of Eq.~\eqref{eq:g_dPsi}}

We start by introducing an identity that we want to apply in the following,
\begin{align} \label{eq:del_identity}
0 = \partial_{\nu,\vq} \braket{\psi_{m,\vk+\vq}}{\psi_{n,\vk}} = \left(\partial_{\nu,\vq} \bra{\psi_{m,\vk+\vq}}\right) \ket{\psi_{m,\vk+\vq}} + \bra{\psi_{m,\vk+\vq}} \left( \partial_{\nu,\vq} \ket{\psi_{n,\vk}}\right) \nonumber\\
\Rightarrow \left(\partial_{\nu,\vq} \bra{\psi_{m,\vk+\vq}}\right) \ket{\psi_{m,\vk+\vq}} = -\bra{\psi_{m,\vk+\vq}} \left( \partial_{\nu,\vq} \ket{\psi_{n,\vk}}\right).
\end{align}

From the stationary Schrödinger (or Kohn-Sham) equation we obtain,
\begin{align}
\bra{\psi_{m,\vk+\vq}(\{\vecr\})} H(\{\vecr\}) \ket{\psi_{n,\vk}(\{\vecr\})} = \delta_{mn} \delta_{\vk, \vk+\vq} \epsilon_{n\vk}(\{\vecr\}),
\end{align}
where we explicitly denoted the dependencies to nuclei positions $\{\vecr\}$.

Now, we applying the derivative $\partial_{\nu,\vq} = \sum_{\kappa\alpha} \frac{1}{\sqrt{M_\kappa}}\ve_{\kappa\alpha,\nu \vq} e^{i\vq \vR_{\kappa}} \frac{\partial}{\partial \vecr_{\kappa\alpha}}$ on both sites and use that bra- and ket-vectors are eigen-states of the Hamiltonian at equilibrium positions,
\begin{align}
\delta_{mn} \delta_{\vk, \vk+\vq} \partial_{\nu,\vq} \epsilon_{n\vk}(\{\vecr\}) &= \partial_{\nu,\vq} \bra{\psi_{m,\vk+\vq}(\{\vecr\})} H(\{\vecr\}) \ket{\psi_{n,\vk}(\{\vecr\})}
\nonumber\\
= \left(\partial_{\nu,\vq} \bra{\psi_{m,\vk+\vq}(\{\vecr\})} \right) H(\{\vecr\}) \ket{\psi_{n,\vk}(\{\vecr\})} +&
\bra{\psi_{m,\vk+\vq}(\{\vecr\})} \left( \partial_{\nu,\vq} H(\{\vecr\}) \right) \ket{\psi_{n,\vk}(\{\vecr\})} +\nonumber\\
\bra{\psi_{m,\vk+\vq}(\{\vecr\})} H(\{\vecr\}) \left( \partial_{\nu,\vq} \ket{\psi_{n,\vk}(\{\vecr\})} \right)
\nonumber\\
= \epsilon_{n\vk}(\{\vecr\}) \left(\partial_{\nu,\vq} \bra{\psi_{m,\vk+\vq}(\{\vecr\})} \right) \ket{\psi_{n,\vk}(\{\vecr\})} +&
\bra{\psi_{m,\vk+\vq}(\{\vecr\})} \left( \partial_{\nu,\vq} H(\{\vecr\}) \right) \ket{\psi_{n,\vk}(\{\vecr\})} +\nonumber\\
\epsilon_{m,\vk+\vq}(\{\vecr\}) \bra{\psi_{m,\vk+\vq}(\{\vecr\})} \left( \partial_{\nu,\vq} \ket{\psi_{n,\vk}(\{\vecr\})} \right).
\end{align}
Now, we can use Eq.~\eqref{eq:del_identity} and obtain,
\begin{align}
&\bra{\psi_{m,\vk+\vq}(\{\vecr\})} \left( \partial_{\nu,\vq} H(\{\vecr\}) \right) \ket{\psi_{n,\vk}(\{\vecr\})} =\nonumber\\
&\delta_{mn} \delta_{\vk, \vk+\vq} \partial_{\nu,\vq} \epsilon_{n\vk}(\{\vecr\}) + \left[ \epsilon_{n\vk}(\{\vecr\})- \epsilon_{m,\vk+\vq}(\{\vecr\}) \right] \bra{\psi_{m,\vk+\vq}(\{\vecr\})} \left( \partial_{\nu,\vq} \ket{\psi_{n,\vk}(\{\vecr\})} \right)
\end{align}
By plugging this into Eq.~\eqref{eq:g_dH_bloch} we obtain Eq.~\eqref{eq:g_dPsi}.

\section{ACKNOWLEDGEMENTS}
We would like to thank the Deutsche Forschungsgemeinschaft
for financial support [projects 511287670, 541495916 and the
Cluster of Excellence e-conversion (Grant No. EXC 2089/1-
390776260)]. Grants for computer time from the Leibniz Supercomputing Centre in Garching are gratefully acknowledged. We further gratefully acknowledge the computing time made available on the high-performance computer Barnard at the NHR Center TUD-ZIH. This center is jointly supported by the Federal Ministry of Education and Research and the state governments participating in the National High-Performance Computing (NHR) joint funding program (http://www.nhr-verein.de/en/our-partners).

\section{COMPETING INTERESTS}
There are no competing interests to declare.


\end{document}